\documentclass[10pt,a4paper]{article}

\usepackage{dcolumn}
\usepackage{bm}
\usepackage{graphicx}
\usepackage{graphicx} 

\begin{document}

\title{Modeling stochastic gene expression under repression}

\author{G.C.P.\ Innocentini, J.E.M.\ Hornos\thanks{work supported by
    FAPESP and CNPq, Brazil}~}

\date{\small
      Instituto de F\'{\i}sica de S\~ao Carlos,
      Universidade de S\~ao Paulo, \\
      Caixa Postal 369,
      BR--13560-970~ S\~ao Carlos, S.P., Brazil}

\footnotetext[2]{\emph{E-mail address:} \textsf{hornos@if.sc.usp.br}}

\maketitle

\setcounter{footnote}{0}
\thispagestyle{empty}
\vspace*{-7mm}
\begin{abstract}

Intrinsic transcriptional noise induced by operator fluctuations is
investigated with a simple spin like stochastic model. The effects of
transcriptional fluctuations in protein synthesis is probed by coupling
transcription and translation by an amplificative interaction. In the
presence of repression a new term contributes to the noise which
depends on the rate of mRNA production. If the switching time is small
compared with the mRNA life time the noise  is also small. In general
the  dumping of protein production by a repressive agent occurs
linearly but  the fluctuations can show a maxima at intermediate
repression. Thediscrepancy between the switching time, the mRNA degradation and
protein degradation is crucial for the repressive control in
translation  without large fluctuations. The noise profiles obtained here 
are in quantitative agreement with recent experiments.

\end{abstract}
\maketitle
\setcounter{page}{1}

        \begin{section}{\large{Introduction}}

The remarkable simplicity of the relation between genetic information
and protein synthesis in the ribosomes, resulting in a symmetrical and
almost universal  genetic code, is the result of a complex process
involving a large number of chemical reactions. The gene transcription
is  assisted by the enzymatic action of regulatory proteins which can
enhance or repress the production of mRNA molecules. The amino acid
assembling  in the ribosome ends the cycle and proteins are produced
and  folded  to a functional form. The pioneer work on lambda phage (1)
have shown the crucial role of regulatory proteins in the control of
such genetic networks (2) and intense experimental and theoretical
investigation (3-13) have been dedicated to the understanding of this
web of interactions.

A deterministic description of the whole process is hampered by the
presence of a frequently small number of molecules in the cell and a
stochastic approach is in general unavoidable (14-15). 
Several efforts have been directed to the numerical simulations of the
genetic networks by considering the full set of chemical
reactions (13-16). Alternatively one can write down a set of dynamical
equations for protein concentrations followed by the inclusion of
fluctuation by the Langevin mechanism (17-21).

A complete description of the cellular control mechanisms requires the
understanding of the complex system of gene interactions and
alternative  efforts have been directed to the identification of
elementary  mechanisms to understand the random aspects of regulatory
networks. A common feature of these models is the use of the rates of
each  chemical reaction to construct Markov process. A first class of
models  can be obtained by considering a single Poissonian model for
transcription to be coupled to stochastic translation (23). The role
of  intrinsic noise can be analyzed in this framework by assuming an
amplifier  interaction designed to reproduced the observed protein
bursts  in translation (23-24). Alternatively regulation can
be  investigated by considering an unique effective stochastic model
for  the combined transcription-translation processes. In this context
repression arises naturally by considering a spin like (27-28)  or
binary  system (26) in the form of a switch. The advantage of spin
like  models is the possibility to use the well established many body
techniques (29). Recently exact solutions for the Master equations of
these system have been obtained in the case of an self interacting
gene(30).  In this article we combine the two strategies by adopting a
two state birth and dead coupled model for the gene, in which the up state
describes the free transcription of DNA while the down state
represents  repressed transcription, to be combined to a second Marcov
process for translation. In the absence of repression our calculations
reproduces the model presented in reference (23). The regulation by an
repressive environmental agent or protein concentration of another
gene is fully considered in the model.

The transcription probabilities obey the equations

        \begin{equation}\label{equa1}          
   \frac{d\alpha_{n}}{dt}=k[\alpha_{n-1}-\alpha_{n}]+ \rho[(n+1)\alpha_{n+1}-n \alpha_{n}]-h\alpha_{n} +f\beta_{n}
  \end{equation} 

  \begin{equation}\label{equa2}          
   \frac{d\beta_{n}}{dt}=k \chi [\beta_{n-1}-\beta_{n}]+\rho[(n+1)\beta_{n+1}-n \beta_{n}]+h\alpha_{n} -f\beta_{n}
  \end{equation}

The stochastic variable n describes, here, the number of mRNA
molecules in the cell. The unrepressed probability for RNA
transcription is $\alpha$(t,n) and $\beta$(t,n) represented the
partially dumped state of the switch. The RNA degradation rate in the
cell is $\rho$ with a life time T=1/$\rho$. The parameter $k$ and
$\chi k$ are the free and repressed transcription rate, respectively.
The parameter h and f controls the repressing efficiency of the
switch. The capability of a protein to bind the repressor site is
coded in the variable h. It is proportional to average repressing
protein number. The unbinding rate reducing the repressing efficiency is f.

The effects of fluctuations on the mRNA population in the protein
dynamics will be investigated in section VI by coupling this birth and
dead process to a stochastic variable describing protein
concentration. The generalization of this approach to a gene network
requires the introduction of more random variables and the
corresponding interactions.
  
This article is organized as its follows. In section II we solve the
model by the introduction of generator functions. A single second
order differential equation for the on state is obtained and reduced
to the canonical form. In section III the physical content of the
model is exhibit by the introduction of the relevant parameters
describing the properties of the gene and environment. Simple
expressions for the distributions, mean values and fluctuations are
also presented. The induced noise is studied in section IV as a
function of the switching time, transcription efficiency and
repressing conditions. Microscopic functions are presented in section
V. In section VI we couple the transcription to translation by the
amplificative interaction of reference (23). The last
section is dedicated to conclusions.
      
        \end{section}
       
        \begin{section}{\large{Master equations for repressed transcription}}
  
The discrete recursive equations implied in the master equations can
be bypassed by the use of generating functions defined by (van Kampen, 1992)              
       
        \begin{equation}\label{equa3}
   \alpha(z,t)= \sum_{n=0} ^\infty{\alpha_{n}(t)z^{n}} \hspace{2cm}\beta(z,t)= \sum_{n=0} ^\infty{\beta_{n}(t)z^{n}}    
  \end{equation} 

The resulting master partial differential equations are
  
  \begin{equation}\label{equa4}
   \frac{\partial \alpha(z,t)}{\partial t}=(z-1)[k\alpha(z,t)-\rho\frac{\partial \alpha(z,t)}{\partial z}]-h \alpha(z,t)+f \beta(z,t)         
  \end{equation}       
      
  \begin{equation}\label{equa5}
   \frac{\partial \beta(z,t)}{\partial t}= (z-1)[\chi
   k\beta(z,t)-\rho\frac{\partial \beta(z,t)}{\partial z}]+h \alpha(z,t)-f \beta(z,t)         \end{equation} 

which must be solved under the condition $\alpha(1,t)+\beta(1,t)=1$ to
ensure probability conservation. The moments can be easily obtained by
the evaluation of this functions and its higher derivatives at
z=1. The full information about the system  requires the calculation
of the distribution functions by the evaluation of derivatives at z=0:  

        \begin{equation}\label{equa6}
   \alpha_{n}(t)= \frac{1}{{n}!}\frac{d^{n}}{dz^{n}}\alpha(z,t)\hspace{1cm}   \beta_{n}(t)=\frac{1}{{n}!}\frac{d^{n}}{dz^{n}}\beta(z,t)
  \end{equation}    

This system is completely integrable due to the existence of Lie
symmetries and the time dependence can also be obtained analytically.
Here we focus our attention in the stationary properties of the system

        \begin{equation}\label{equa7}
   (z-1)(k \alpha_{S}+\rho\frac{d\alpha_{S}}{dz})-h \alpha_{S}+f\beta_{S}=0
  \end{equation} 

  \begin{equation}\label{equa8}
   (z-1)(\chi k\beta_{S}+\rho \frac{d\beta_{S}}{dz})+h\alpha_{S}-f\beta_{S}=0
  \end{equation} 

The coupled equation can be transformed in a single second order
differential equation by replacing      
       
        \begin{equation}\label{equa9}
   \beta_{S}=\frac{1}{f} \left[\rho(z-1)\frac{d\alpha_{S}}{dz}-(zk-k-h)\alpha_{S}\right]
  \end{equation}

in equation (8). The resulting equation in the canonical form is       
      
        \begin{equation}\label{equa10}
   \frac{d^2 \alpha_{S}}{dz^2}+p\frac{d \alpha_{S}}{dz}+q\alpha_{S}
  \end{equation}

where the functions p and q are given by 
     
        \begin{equation}\label{equa11}
   p= \frac{(1-z)(1+\chi)k+\rho+h+f}{\rho(z-1)}
  \end{equation}
      
  \begin{equation}\label{equa12}
   q=\frac{k^2 {\chi}(z-1)-k(\rho+f+{\chi}h)}{\rho^2(z-1)}
  \end{equation}

The simple poles at z=1 and the asymptotic irregularity suggests
solutions in terms of the hypergeometric functions. In fact
introducing the variable 
  
  \begin{equation}\label{equa13}
   \eta= \frac{k (\chi-1) (z-1)} {\rho}
  \end{equation}      

and performing the transformation

        \begin{equation}\label{equa14}
   \alpha_{S}= exp\hspace{0.1cm}(\frac{k(z-1)} {\rho})\tilde{\alpha}(\eta)
  \end{equation}

we obtain the Kummer equation       

        \begin{equation}\label{equa15}
   \eta \frac {d^{2}\tilde{\alpha}}{d{\eta}^{2}} +( b+1-\eta) {\frac {d\tilde{\alpha} }{d\eta}} -a \tilde{\alpha} = 0 
  \end{equation} 
  
with parameters

        \begin{equation}\label{equa16}
   a= \frac {h} {\rho}
   \end{equation} 
  
  \begin{equation} \label{equa17}
   b=  \frac {h+f} {\rho}
  \end{equation} 
       
The regular solution at z=1 ($\eta=0$) can be written in terms of the
Kummer M-functions    
     
   \begin{equation}\label{equa18}
    \alpha_{S}= \frac{b-a} {b} exp\left(\frac{k(z-1)} {\rho}\right) M(a,b+1;\eta)
   \end{equation} 
   
resulting in an analogous equation for $\beta$ 

         \begin{equation}\label{equa19}
    \beta_{S}= \frac{a}{b}exp\left(\frac{k(z-1)}{\rho}\right)M(a+1,b+1;\eta)  
   \end{equation}

A simple formula can also be obtained for the total probabilities
       
        \begin{equation}\label{equa20}
   \phi_{S}= exp\left(\frac{k(z-1)}{\rho}\right)M(a,b;\eta)  
  \end{equation}

The advantage of the use of generating functions is the analytical
elimination of the singular solution (a U-Kummer function). In
numerical calculations the presence of this component is the source of
numerical instabilities in the algorithm
       
        \end{section}
     
  \begin{section}{\large{The biological content of the model}}

The relevant information about the properties of the switch in the
proteomic environment is specified by two time scales
        
  \begin{equation}\label{equa21}
   T= \frac{1}{\rho}\hspace{1cm} \tau=\frac{1}{h+f} 
  \end{equation}

The mean life of the mRNA molecules in the cells is coded in $T$, it
is typically of order of minutes. The meaning of the switching time
$\tau$ is related to the time evolution of the stochastic dynamics
process. In fact, a simple equations for the probabilities
$p_\alpha(t)$ and $p_\beta(t)$ to find the gene free or repressed by a
protein in the operator site is obtained by setting z=1 in equations
(1) and (2):
    
        \begin{equation}\label{equa22}
   \frac{dp_{\alpha}(t)}{dt} = -h p_{\alpha}(t)+f p_{\beta}(t)         
  \end{equation}       
      
  \begin{equation}\label{equa23}
   \frac{dp_{\beta}(t)}{dt} = +h p_{\beta}(t)-f p_{\beta}(t)         
  \end{equation} 

summing Eq.2 over n with the use of probability conservation. The solution is:
          
        \begin{equation}\label{equa24}
   p_\beta(t)= \frac{h}{h+f}+e^{-(h+f)\;t/\rho}(p_\beta(0)-\frac{h}{h+f})
  \end{equation}

The switching time is a intrinsic property of the gene. It doesn't
depends on the birth and dead constants, $k$ and $\rho$. The
adimensional parameter
      
   \begin{equation}\label{equa25}
   \epsilon= \frac{h+f}{\rho}
  \end{equation}

contains the information about the flexibility of the gene
switch. High values of $\epsilon$ corresponds to fast switches
reaching the equilibrium configuration before the gene transcription
is stationary. As we show above the average number of mRNA copies doesn't
depends on the switching time but the noise is strongly affected by
this parameter. 
            
The influence of the environment on the gene transcription is measured
by the asymptotic probability to find the gene repressed 

        \begin{equation}\label{equa26}
   p_\beta= \frac{h}{h+f}\hspace{1cm}p_\alpha=\frac{f}{h+f}
  \end{equation}

In the limit $p_\beta=0$ we recover the model presented in
(23) with only one transcription mode and a Poissonian
distribution where $N = \frac{k}{\rho}$ is the average number of mRNA
copies. Highly repressed genes belongs to the $p_\beta=1$ sector where
the mRNA transcription is suppressed or occurs in the repressed mode
($\chi \neq 0$). The parameter $\chi$ takes in account the possibility of
a small rate of transcription in the (partially) repressed mode. In
summary, N measures the efficiency of the
transcription. A gene with a large value of N produce mRNA in
abundance, in the absence of repression, but also can be damped by
regulatory proteins or an external factor to $\chi N$ particles in
average. Inefficient transcriptions as occurs in certain genes of
E.Coli, for example, corresponds to N around one or two, while in
copious transcription N increases an order of magnitude.  The effects
of the protein concentration or another repressing factor is controlled by
$p_\beta$ which can be phenomenologically related to protein
concentration by a Hill function, for example. The relative switch
frequency scales the capability of the gene to flip from an arbitrary
condition to it's asymptotic value. The relevant features of the
probabilities can be obtained by inspecting the generator functions
written in terms of these parameters:
        
        \begin{equation}\label{equa27}
   \alpha_{S}=p_{\alpha} exp\left(N(z-1)\right) M(\epsilon p_{\beta},\epsilon+1; N(z-1) (\chi-1))
  \end{equation} 

for the up state

        \begin{equation}\label{equa28}
   \beta_{S}=p_{\beta} exp\left(N(z-1)\right) M(\epsilon p_{\beta}+1,\epsilon+1;N(z-1) (\chi-1))
  \end{equation} 

for the down state

        \begin{equation}\label{equa29}
   \phi_{S}=exp\left(N(z-1)\right) M(\epsilon p_{\beta},\epsilon;N(z-1)(\chi-1))
  \end{equation} 

for the total probability

In the absence of a repressive agent $p_{\beta}=0$ and $p_{\alpha}=1$,
the Kummer function reduces to one and we have a Poissonian process
producing N mRNA copies in average. In general the generator function
is a power series on N, modulated by $\chi$ with coefficients
depending of the repression parameter and switching time via a
Pochhammer function. If the gene interact with a stochastic factor, as
in the case of fluctuating proteins due to another gene, the
parameters N, $p_{\beta}$ and eventually $\epsilon$ should be
considered random functions of the corresponding proteomic field. In
the model we have two different sources for hampering or enhancing
transcription. The transcription
efficiency caused by the availability of polymerases affects the free
mean number N while the density of repressive proteins modifies $p_{\beta}$. 

If we introduce a power series in the above equation we obtain
recursion relations for a second order Markov process for the total
probability. The probability is obtained taking successive derivatives
of the generator function by the use of the Leibnitz rule and the
formulas for the derivatives of Kummer functions:
       
        \begin{equation}\label{equa30}        
   \alpha_{n} = p_{\alpha}(N)^n \frac {exp(-N)} {n!} \sum_{s=0}^{n}
   {n\choose s} (\chi-1) ^{s}\frac{(p_{\beta}\epsilon)_s} {(\epsilon+1)_s} M(p_{\beta}\epsilon+s,\epsilon+1+s;N(1-\chi)) 
        \end{equation}
       
        \begin{equation}\label{equa31}        
   \beta_{n} = p_{\beta}(N)^n \frac {exp(-N)} {n!} \sum_{s=0}^{n}
   {n\choose s} (\chi-1) ^{s}\frac{(p_{\beta}\epsilon+1)_s} {(\epsilon+1)_s} M(p_{\beta}\epsilon+1+s,\epsilon+1+s;N(1-\chi)) 
        \end{equation}
       
        \begin{equation}\label{equa32}        
   \phi_{n} = (N)^n \frac {exp(-N)} {n!} \sum_{s=0}^{n}
   {n\choose s} (\chi-1) ^{s}\frac{(p_{\beta}\epsilon)_s} {(\epsilon)_s} M(p_{\beta}\epsilon+s,\epsilon+s;N(1-\chi)) 
        \end{equation} 
 
The Poisson distribution appears with an envelope function composed by a
superposition of distributions implementing the repression.
In figure (1) we show a typical  distribution for the efficient
transcription of  gene producing N=40 mRNA copies. The repressed mode
produces $20$ of the free mode ($\chi=1/5$). The gene is slow
$\epsilon=1/2$.
The $\alpha$ and $\beta$ components are shown together with the sum in
the figure. The mean peak corresponds to the free production and the
secondary to the repressed transcription.

      \end{section}

      \begin{section}{\large{Mean values and noise}}

The closed form for the generator functions and the friendly
properties of the Kummer functions allows one to obtain 
simple formulas for all the distribution moments evaluating the
derivatives at the point z=1. The mRNA mean value produced in both
models in the cell is

      \begin{equation}\label{equa33}
       \langle n\rangle= N(p_\alpha+\chi p_\beta)
      \end{equation}

It is linear in the gene efficiency N  and in the repression parameter
$p_{\beta}$. The switch life time do not appear in the mean
value. If we assume that the probability to find the gene in the on
mode without a binding repressing protein is related to the
concentration by a Hill function

      \begin{equation}\label{equa34}
       p_{\alpha}= \frac{1}{1+(<m>/K)^l}
      \end{equation}

where K is the threshold concentration of the repressor
agent and l is steepness factor, we obtain the curves shown
in Fig.2. The higher curve corresponds to a gene that can
be transcripted poorly even in the presence of the
repressor agent ($\chi=1/5$). The free transcription is not
affected by low concentrations but will decay abruptly
after the threshold concentration when the mean values
reaches it's repressed value. The second curve describes a
$\chi=0$ switch and after threshold there is no
transcription in the cell.

The possibility of repressed production causing the appearance
of two peaks statistical distributions requires the use of the
moments of the on and off components. In this case the total
mean value and also the total fluctuation will have a limited
meaning. However simple formulas can also be obtained for the
partial mean values:

        \begin{equation}\label{equa35}
         \langle n\rangle_\alpha = \langle n\rangle +\frac{N(1-\chi)p_\beta}{1+\epsilon}
         \end{equation}
         \begin{equation}\label{equa36}
          \langle n\rangle_\beta = \langle n\rangle -\frac{N(1-\chi)p_\alpha}{1+\epsilon}
         \end{equation}
         
The effects of the transcription efficiency and the
switching frequency in the noise can be evaluated from the
standard deviation
         
         \begin{equation}\label{equa37}
         \langle n^2\rangle-\langle n\rangle^2 = \langle n \rangle +
         \frac{N^2(1-\chi)^2p_\alpha p_\beta}{1+\epsilon}
         \end{equation}
         
 In both limits $p_\beta =0$, absence of repression, and
 $p_{\beta}=1$, total repression, the stochastic process is
 Poissonian. The square root deviation as a function of the repressing
 parameter $p_\beta$ is shown in Fig.3 for the copious transcription
 of a gene. The curves correspond to several values of the
 switching frequency. For slow genes (small $\epsilon$) the noise
 increases with the repressing
 parameter until maximal fluctuations are reached for 
         
         \begin{equation}\label{equa38}
         (p_\beta)_{max} =\frac{1}{2}-\frac{1}{2}\frac{1+\epsilon}{N(1-\chi)}
         \end{equation}

After this limit the fluctuations will be smaller until a minimal value
for total repression. In this case, to reach the stationary
configuration, the gene will intercalate on and off
configurations many time causing strong noisy in the copious
transcription. We are describing the noise in terms of the mean
deviation instead of the Fano factor that is exhaustive used in
physics. The two alternatives are of course equivalent but the Fano
factor measures the deviation from a Poisson distribution and don't
have the intuitive appeal of the distribution width. For large values
of $\epsilon$ corresponding to fast switches the parabolic fluctuations
will not reach the maxima and decreases monotonically. The repression
drops the mean value of the mRNA transcription and the fluctuations
simultaneously. Inspecting equation (37) we see the presence of a
factor $\epsilon+1$ in the denominator responsible for the noise
attenuation. The disparity between the mRNA life time in the cell and
the binding-unbinding characteristic time leading to large values of
$\epsilon$ ensures low fluctuations and silent transcription.
In Fig.4 the fluctuations of a poorly
transcribed gene are shown. The enhancement of the noise is limited
by the presence of few mRNA copies in the cell and
suppressed totally for fast switches. The critical switch time is

       \begin{equation}\label{equa39}
        (\epsilon)_{crit}= N(1-\chi)-1
       \end{equation}

We should keep in mind that the abundant transcription of a gene
requires the enzymatic action to eliminate the unused messengers
expending energy.
       
     \end{section}

       \begin{section}{\large{Probabilities}}

The description of the random gene dynamics in terms of the
two first moments are satisfactory for local statistical
distributions which is the most common case. The fluctuations
will describe properly the departure from the deterministic
or macroscopic behaviour. However the possibility of repressed
production or the appearance of strongly non local
probabilities requires the full knowledge of the probability
distribution. The solubility of the model allow us to
investigate the behaviour of the probabilities as a function
of the biophysical parameters of the model.

The probabilities as a function of mRNA population for
abundant transcription of a gene is shown in Fig.3. We assume
that without repression N=40 mRNA copies would be produced in
the stationary state. A small repressed production $\chi=1/5$
is also allowed. In Fig.3A the behaviour of a slow switch,
$\epsilon=1/2$, is investigated as a function of the
repressor activity $p_{\beta}$. In the absence of repression
we see a Poissonian peak around n=40. The presence of
moderated repression, $p_{\beta}=0.25$ has three main
effects:reduces the Poissonian maximum, displaced toward
lower population and creates a secondary peak at lower
population.The switch is been turned off by the action of the
repressing factor. The situation is already reversed by
intermediate repression, the secondary maximum now is
dominant indicating that the transcription of the gene occurs only in the off,
or partially repressed state. Under strong repression
$p_{\beta}=0.75$ the Poissonian peak disappears totally and
finally under a full repression the gene is off been
trancripted secondarily at low rates. Increasing the switching
time will favour the delocation of the probability as
shown in Fig.3B. The free and the totally repressed curves
are the same as in Fig.3A but the intermediate curve
$p_{\beta}=0.5$ is widespread over the population. The
situation is even more dramatic if the gene is faster as is
shown in Fig.3C. A plateau with equally probable population
appears under intermediate repression. The mean value and
fluctuation have a limited meaning in this case. In Fig.3C
we can see the disappearance of the double peaks under any
repressive condition. Finally in Fig.3E and Fig.3F the
probabilities recovers the Poissonian shape and moves
adiabatically under the repression action. In Fig.3E as a
solitary wave keeping the shape undisturbed and in the case
of very fast switches (Fig.3F) decreasing the mean deviation
until total repression.  
The turn off of a gene by environmental repression is shown in Fig.4A
for the case of efficient transcription. We choose $\chi=0$ therefore
there is no residual transcription in the off mode. The first
curve, incomplete in the plot, represents the Poissonian unrepressed
distribution, and the other shows the evolution of the probability
for growing repression. A kink is formed for intermediate repression
($p_\beta=0$ centered around $30 < n < 40$ separating the equally
distributed probability region from the null region. The peak
concentrate at n=0 for $p_\beta=1)$ shows the total repression of the
transcription. The inefficient transcription is shown in figures 4B and
4C. The last corresponds to high values of $\epsilon$ and shows the
typical Poissonian behaviour. However in figure 4B we can see the
delocation of the statistical distribution typical of slow
switches.
In the set of figures, 5A, 5B, 5C, we show the tridimensional plot of
the probabilities as a function of $\epsilon$ for null, intermediate and
total repression. At intermediate repression the $\epsilon$ variation
transform the double peak distribution at low $\epsilon$ in a non-local
function until the Poisson like behaviour reappears for fast switches.
The critical value of $\epsilon$ for which the fluctuation decreases
without an enhancement of fluctuation is given by

 In Fig.5 the probability surfaces are show as a function of the
 population and switching time for unrepressed, intermediate and fully
 represses systems. The switching time can change considerably the
 distribution pattern at intermediate repression.      
       \end{section}

       \begin{section}{\large{Induced noise in protein production}}

The effects of the transcription fluctuations on the protein
concentration can be investigate allowing our probabilities to depend
on a new stochastic variable m for the protein concentration. This
will be refrased in the language of generator functions by the
appearance of a new continuous variable y and we will deal with a time
dependent partial differential equation in both variables. Among a
large variety of probability conserving coupled partial equations the 
simplest is given by:

       \begin{equation}\label{equa40}
        \frac{\partial \alpha}{\partial \tau}=(z-1)[N\alpha -
        \frac{\partial \alpha}{\partial z}] - p_{\beta}\epsilon\alpha+
        p_{\alpha}\epsilon \beta + (y-1) [\sigma z\frac{\partial \alpha}{\partial z}-\eta\frac{\partial \alpha}{\partial y}]       
         \end{equation}       
      
        \begin{equation}\label{equa41}
        \frac{\partial \beta}{\partial \tau}=(z-1)[\chi N\beta -
        \frac{\partial \beta}{\partial z}] + p_{\beta}\epsilon\alpha-
        p_{\alpha}\epsilon \beta + (y-1) [\sigma z\frac{\partial \beta}{\partial z}-\eta\frac{\partial \beta}{\partial y}]       
         \end{equation}

The parameters $\sigma$ and $\eta$ are equal to $k_P / \rho$ and
$\rho_P / \rho$ , respectively, where $k_P $ and $ \rho_P$ are the
birth and dead rates of the protein production effective
reaction. This amplifier coupling reproduces the bursts in protein
production that are observed experimentally. For
$p_\beta=0$ the transcription is Poissonian and the usual results
are obtained. However in the general case both
effects, the induced noise caused by the coupling and the repressive
noise are present. Although the full distributions are hard to
obtain, the moments can be recursively obtained by the tradition
techniques (Kampen, 1992). The mean value for protein concentration is
       
       \begin{equation}\label{equa42}
        \langle m \rangle= \frac{\sigma N}{\eta}(\chi
        p_\beta+p_\alpha)
       \end{equation} 

 The adimensional parameter $\eta$ is small, typically around 1/30 or
 less due to the discrepancy between the time scales for transcription
 and translation. The mRNA free mean number N is amplified by a factor
 $\sigma / \eta$. The fluctuations are

       \begin{equation}\label{equa43}
       \langle m^2 \rangle-\langle m \rangle^2 = \langle m \rangle+
       \frac{\sigma^{2} N}{\eta(\eta+1)}\left(p_{\alpha}+\chi
       p_{\beta}+\frac{N p_{\alpha} p_{\beta}(\eta+\epsilon+1)((\chi-1)^2)}{(\eta+\epsilon)(\epsilon+1)}\right)
       \end{equation}

  The terms on parenthesis are caused by the environmental
  repression. The relative frequencies $\epsilon$ and $\eta$ define
  three time scales for the phenomena. The switching time associated to
  the binding-unbiding of proteins to the operator site is the smaller
  time in the problem. A second scale, one order of magnitude higher,
  is defined by the mRNA degradation rate. Finally the protein
  degradation rate is the slower reaction time. Ultimately the slow
  down of the noise in the global process is ensured by the
  discrepancy of these scales and guarantee the precision in cellular
  reproduction. In Fig. 6 we show the standard deviation in three
  cases with the parameters adjusted to give the same protein mean
  concentration in the cell,
  ($\langle m \rangle =3200$) in order to compare the two possible
  strategies. The monotonically decreasing function shows the behaviour
  of the fluctuation for inefficient transcription (N=4). The
  quadratic dependence of the variance on the protein parameter $\sigma$ is
  responsible for the large deviation in the absence of repression. The
  second curve correspond to N around 6 and $\epsilon$=5. The crossing point
  around $p_\beta  \approx$ 0.5 correspond to states with the same mean number
  , fluctuation and repression but with different switch velocities and
  production rates. In the third curve
  fluctuations are enhanced parabolic by low repression until a maxima
  is obtained. Beyond this value the deviation is attenuated until
  the system is off. For intermediate values of $p_\beta$ both process
  are equivalent resulting in comparable mean values and noise. 
  Finally we consider a gene with an increasing activity by an inductor
  A concrete example is a tetracicline repressor under the control of 
  IPTG  in E. Coli. This gene have been used as a 
  part of a synthetic gene cascade in recent experiments. Instead of the mean square 
  root deviation of fano factor we display the noise using the quantity   
      
      \begin{equation}\label{equa44}
       \Delta^{2}= \frac{\langle m^2 \rangle-\langle m \rangle^2}{\langle m \rangle^2}
    \end{equation}  
       
       We relate the model parameter $p_\alpha$ with the repressor concentration by a
       Hill function

     \begin{equation}\label{equa45}
       p_{\alpha}= \frac{(c/k)^{\theta}}{1+(c/k)^{\theta}}
     \end{equation}
       
       where {\bf c} is the concentration an {\bf k} a parameter.
       The protein mean value and the corresponding noise are shown
       in figure 7.A and 7.B and have been obtained replacing $p_\alpha$
       from equation (45) in equations (42) and (43), respectively.
       It is remarkable the agreement between the concentration and the 
       noise profiles with the experimental data of reference (25). In the 
       present case we have single gene, global fluctuation have been discarded 
       and $p_\alpha$ depends  only on the mean concentration of the inductor.
       In figure 8.A and 8.B we show the opposite situation in which the concentration
       decrease and the noise increases. We can see in the figure a maxima in the
       fluctuation for intermediate concentrations. Again the profile is similar
       to experimental results even under the the mentioned differences and restrictions.  
              
       \end{section}  
       
       \begin{section}{\large{Conclusions}}

The hampering of noise in genetic networks by the action of
regulatory proteins is supported by some experiments and also
by the common intuition that repression slow down
fluctuations. However this is not a mandatory rule. The
presence of fluctuations is governed by the transcription
efficiency of the gene, by the repressor concentration but
also by the switching time describing the gene agility to
interchange from a repressed state to a free mode. These
components combined in a simple stochastic model for
repressed transcriptions allow us to establish conditions for
the occurrence of noise in a elementary network component.
        
The transcription of a gene at slow rates is silent even if
the gene switch turn off slowly. The repression agent
decreases the fluctuations in accord to the concentration of
the controlling factor. In the common case of fast switches
the microscopic distributions are Poisson-like and the
flip-flop of the gene configurations induced by repression
will cause small corrections.

In the opposite case when the transcription is copious the
mRNA population fluctuates in the cells strongly in the case
of slow switches. The noise is smaller than in the
unrepressed case only for ultra fast switches revealing the
fundamental role of the time switch in the dynamics of the
noise.

The microscopic probabilities obtained due to the
integrability of the coupled Markovian process shows
non-local, broad bands population distributions under highly
noisy conditions. The possibility of partial repressed
transcription allowed in the model causes the appearance of
double peaks in the probabilities due to the repressed and
free transcription.

The two fundamental mechanisms which causes the decreasing in
transcription rates, the shortage of polymerases and the
presence of a repressor agent affects the fluctuations
differently. In the first case the average mRNA population
decreases homogeneously with the gene efficiency
parameter. In the second case we see the parabolic dependence
with the repression parameter.

The coupling of the transcription to translation by the
amplifying mechanism reproduces the occurrence of protein
bursts under repression with a rich noise structure.

Our model is minimal in the sense that: (i) It is a simple
and soluble second order stochastic model for repressed
transcription.(ii) The introduction of two states for the
gene contemplating the repressed transcription introduces the
two time scales involved in the process. (iii) The effects on
translation proceeds by minimally coupling the transcription to
protein production. (iv) Gene interactions and auto regulation
can be implemented by allowing fluctuations in  the model
parameters.

The evolutionary requirements of noise to enforce
differentiation, prompts the need of experiments, beyond the
poorly transcribed genes, as maltT of E.coli, designed to
probe the noise in copious transcription.

The comparison of our model with experiments shows a good agreement
even considering the the limitations of our single gene treatment to
explain the observed cascade effects under global noise.

Acknowledgment: J.E.M. Hornos wishes to thanks to J.N. Onuchic for the 
introduction into the field, P. Wolynes for helpful discutions and 
W. Arber for the memories on lambda phage experiments.

       \end{section}

      \newpage
      \begin{figure}[!htb]
      \centering
      \begin{minipage}[b]{0.65\linewidth}
      \includegraphics[width=\linewidth,angle=0]{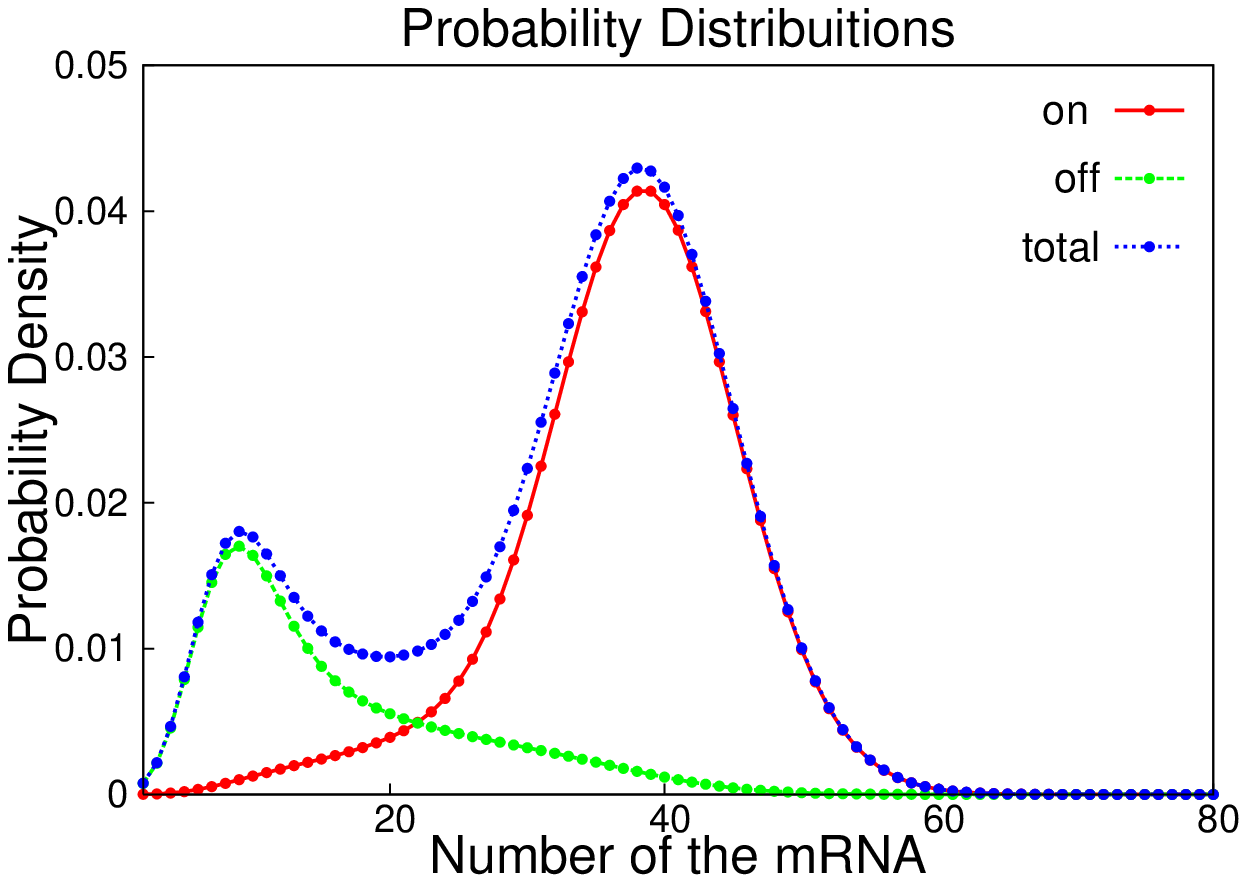}
      \end{minipage}
      \caption{Probabilities for the on, off states are show together
      with the total probability. The parameters are N=40,
      $\chi$=1/5, $\epsilon=1/2$ and $p_\beta=0.20$}
      \end{figure}

       \newpage
       \begin{figure}[!htb]
       \includegraphics[width=3.0in,angle=0]{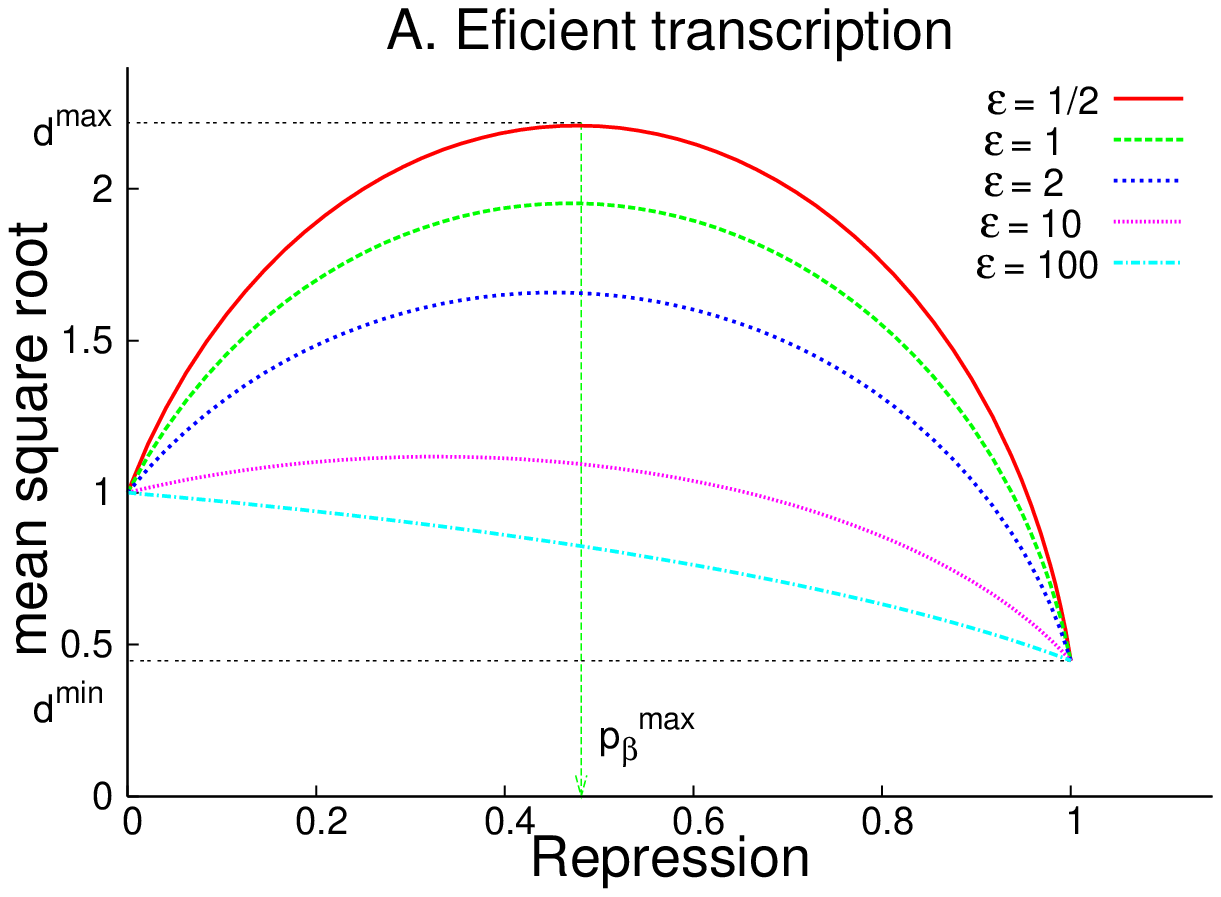}
       \includegraphics[width=3.0in,angle=0]{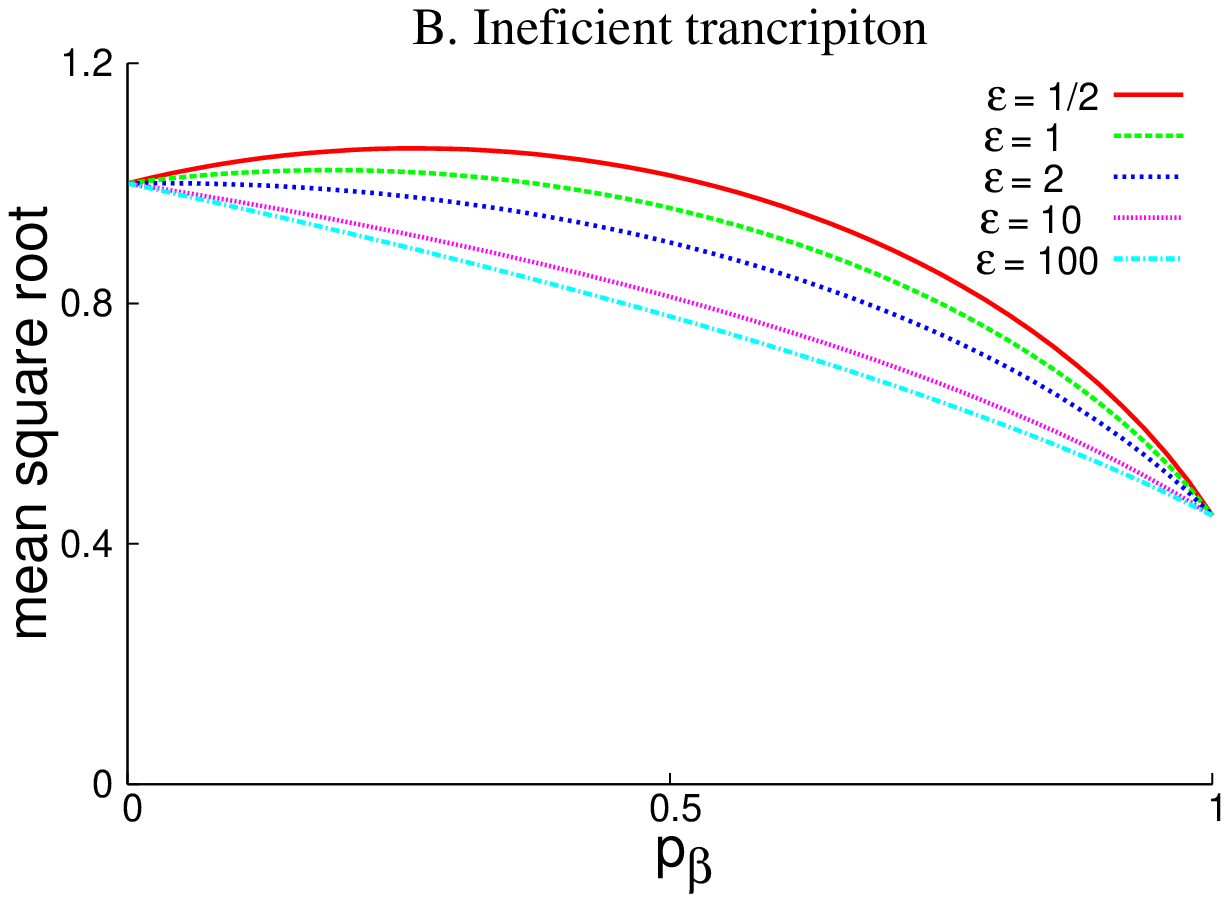}
       \caption{Figures 2A and 2B, standard deviation for high and
       low transcription with parameter N=40 and 4 respectively. Repressed
       production $\chi=1/5$}
       \end{figure}

       \newpage  

       \begin{figure}[!htb]
       \includegraphics[width=3.0in,angle=0]{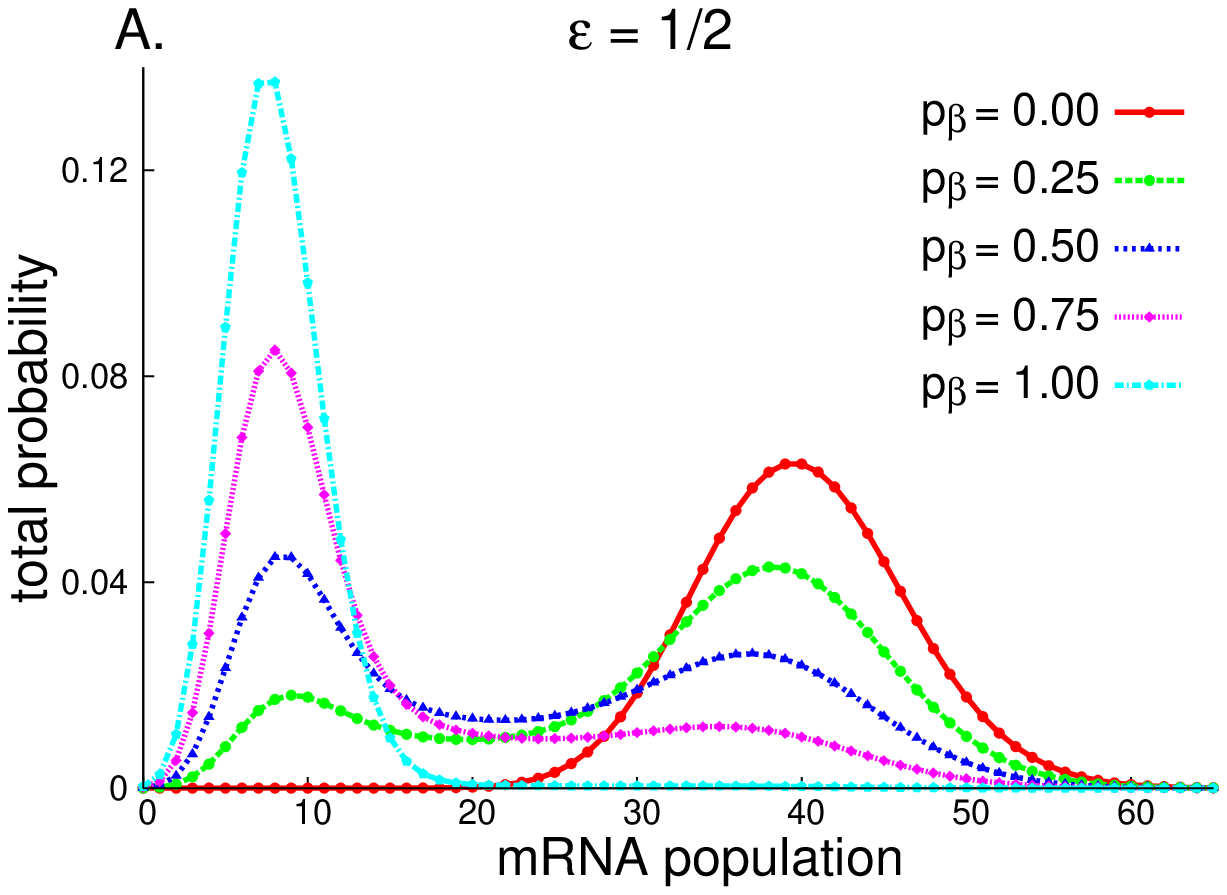}
       \includegraphics[width=3.0in,angle=0]{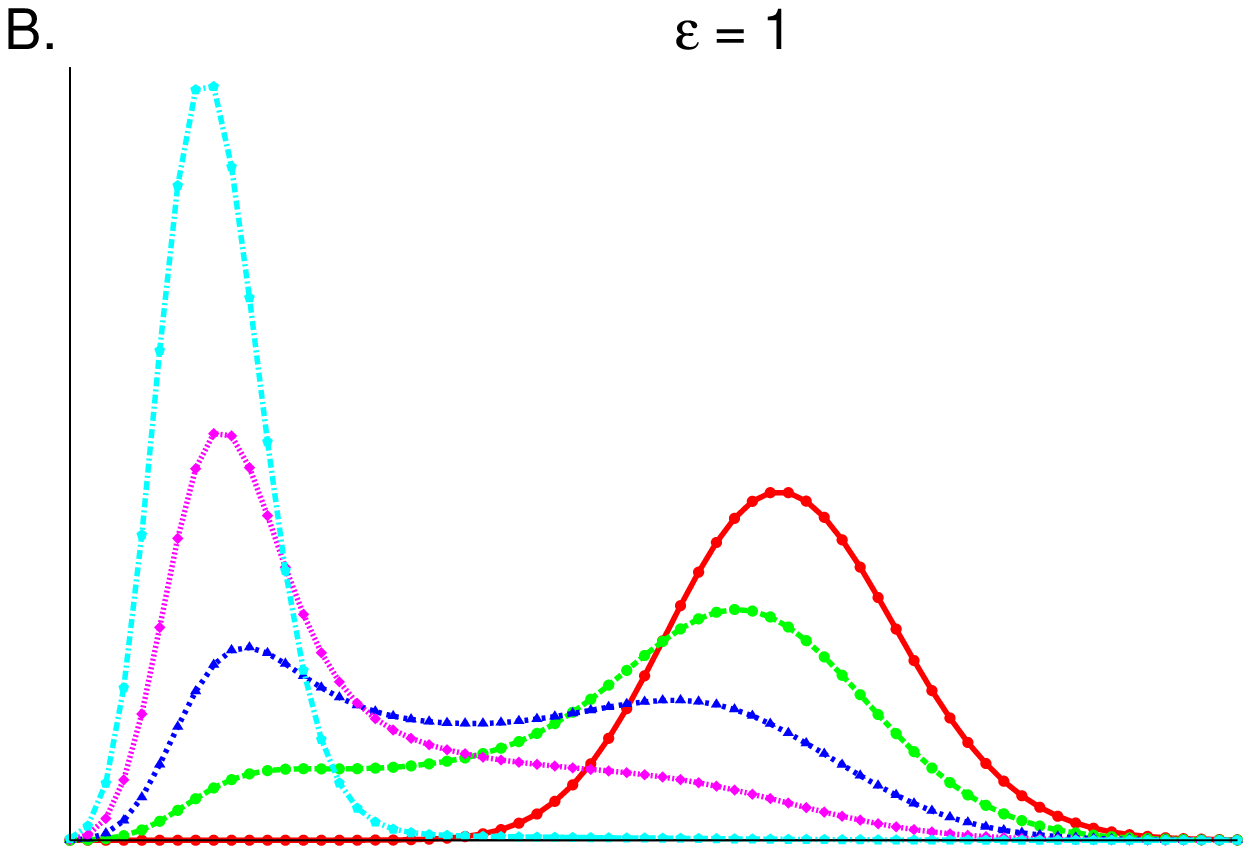}\\
       \includegraphics[width=3.0in,angle=0]{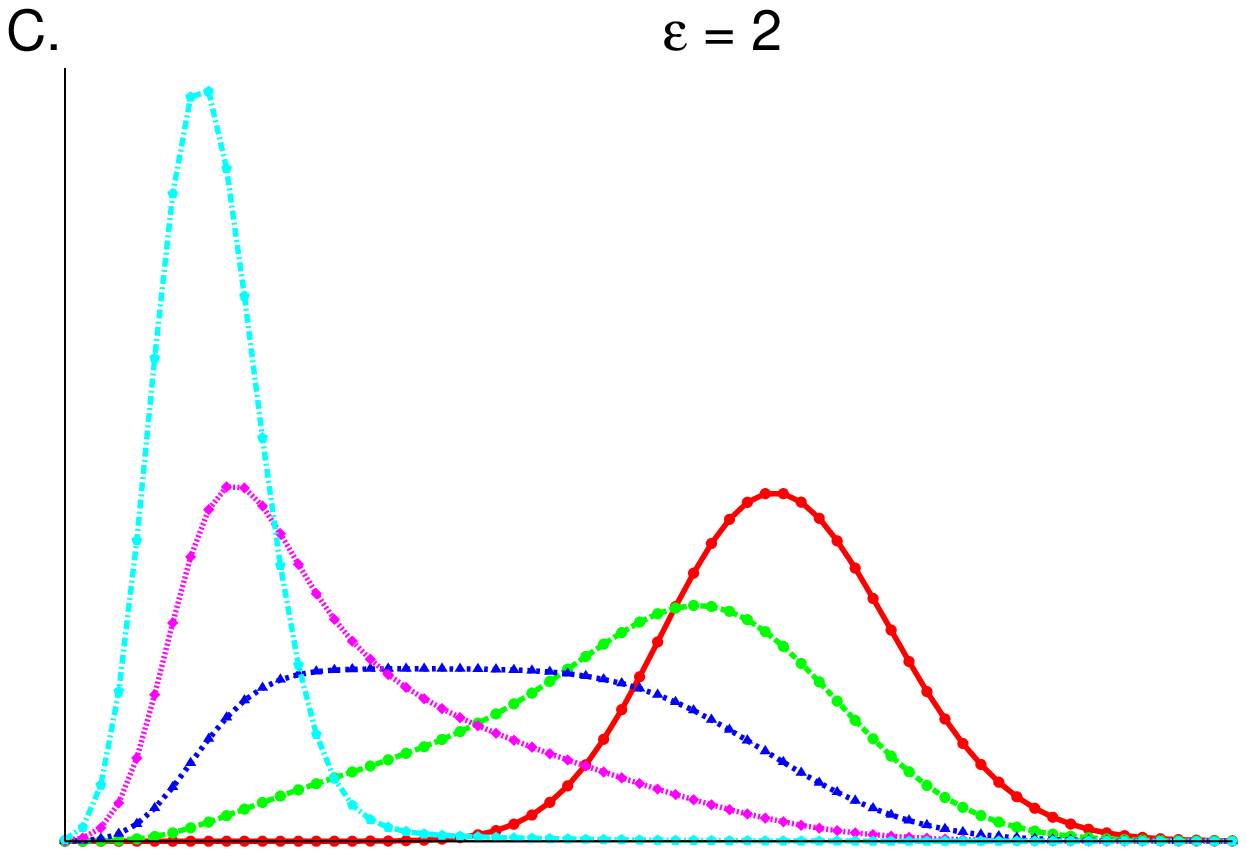}
       \includegraphics[width=3.0in,angle=0]{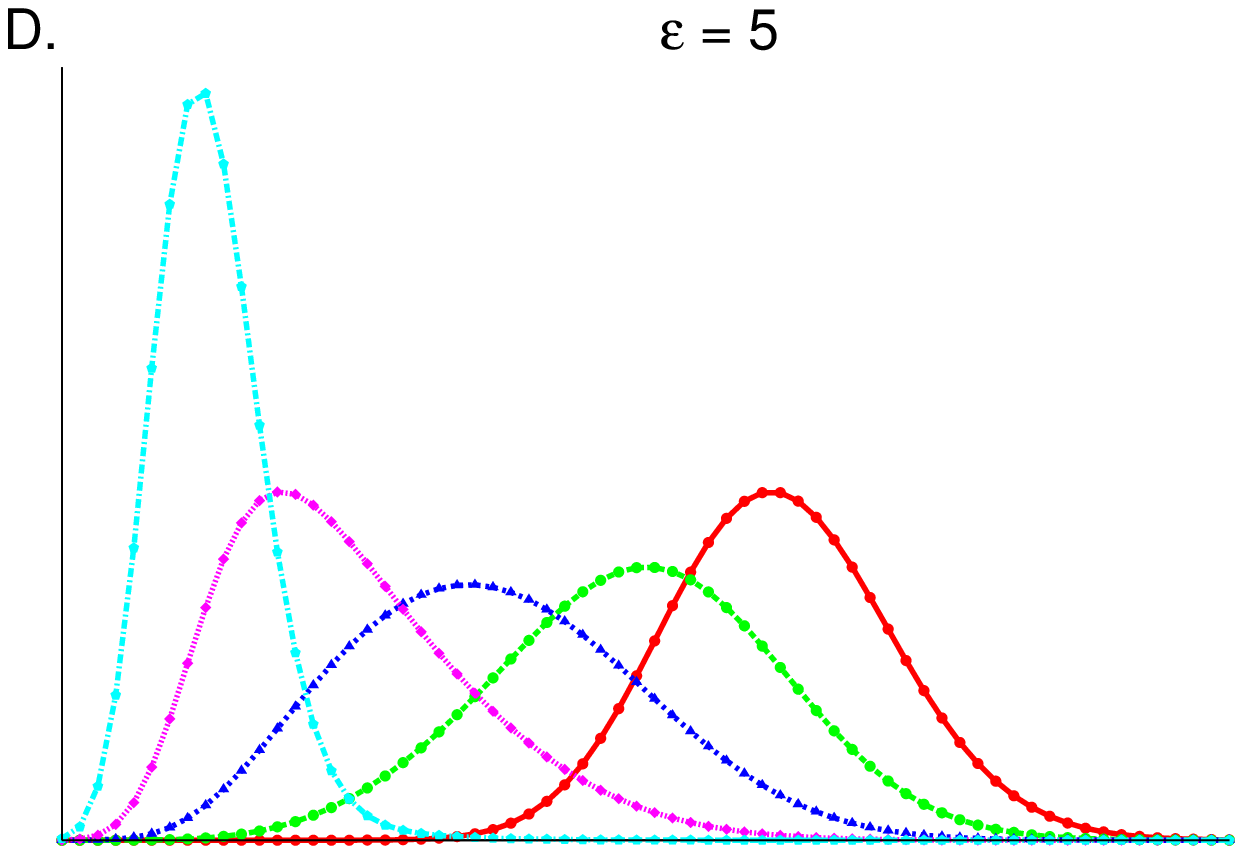}\\
       \includegraphics[width=3.0in,angle=0]{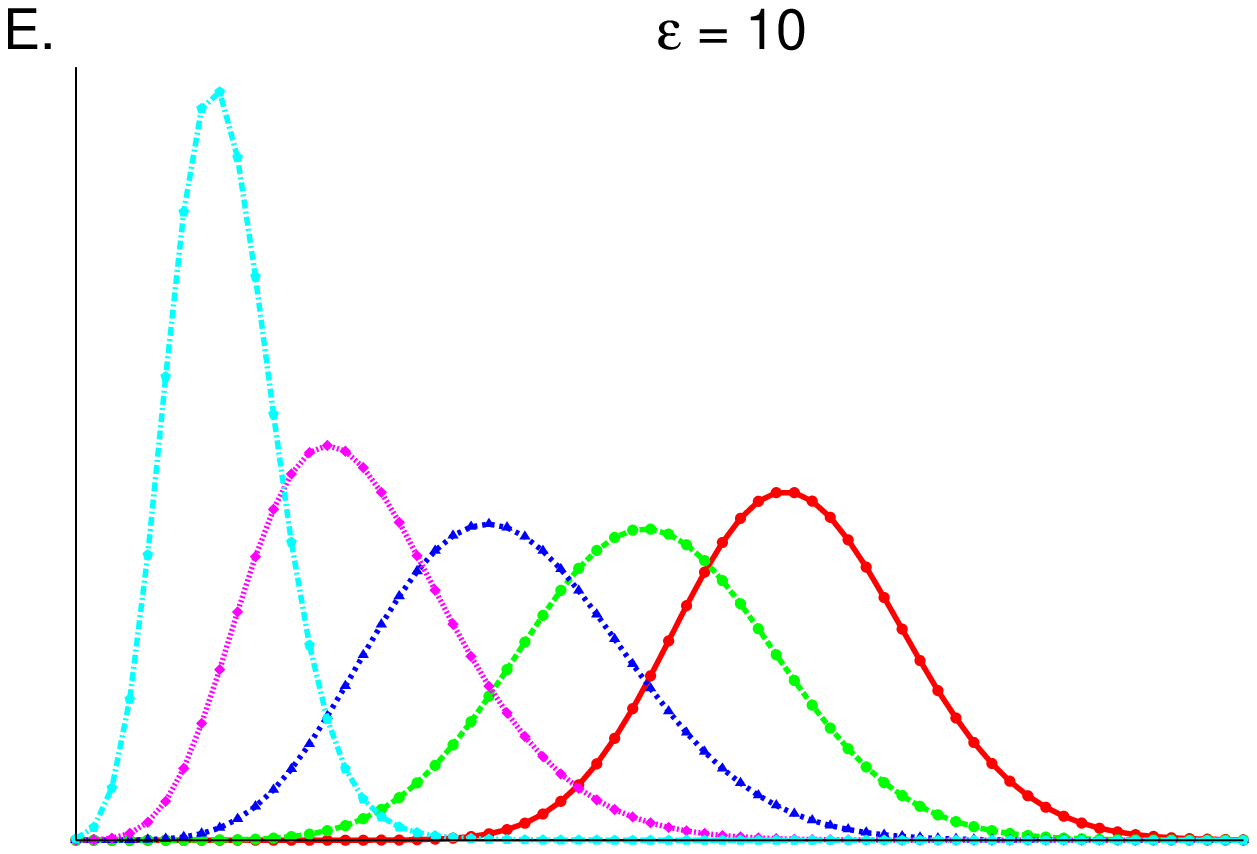}
       \includegraphics[width=3.0in,angle=0]{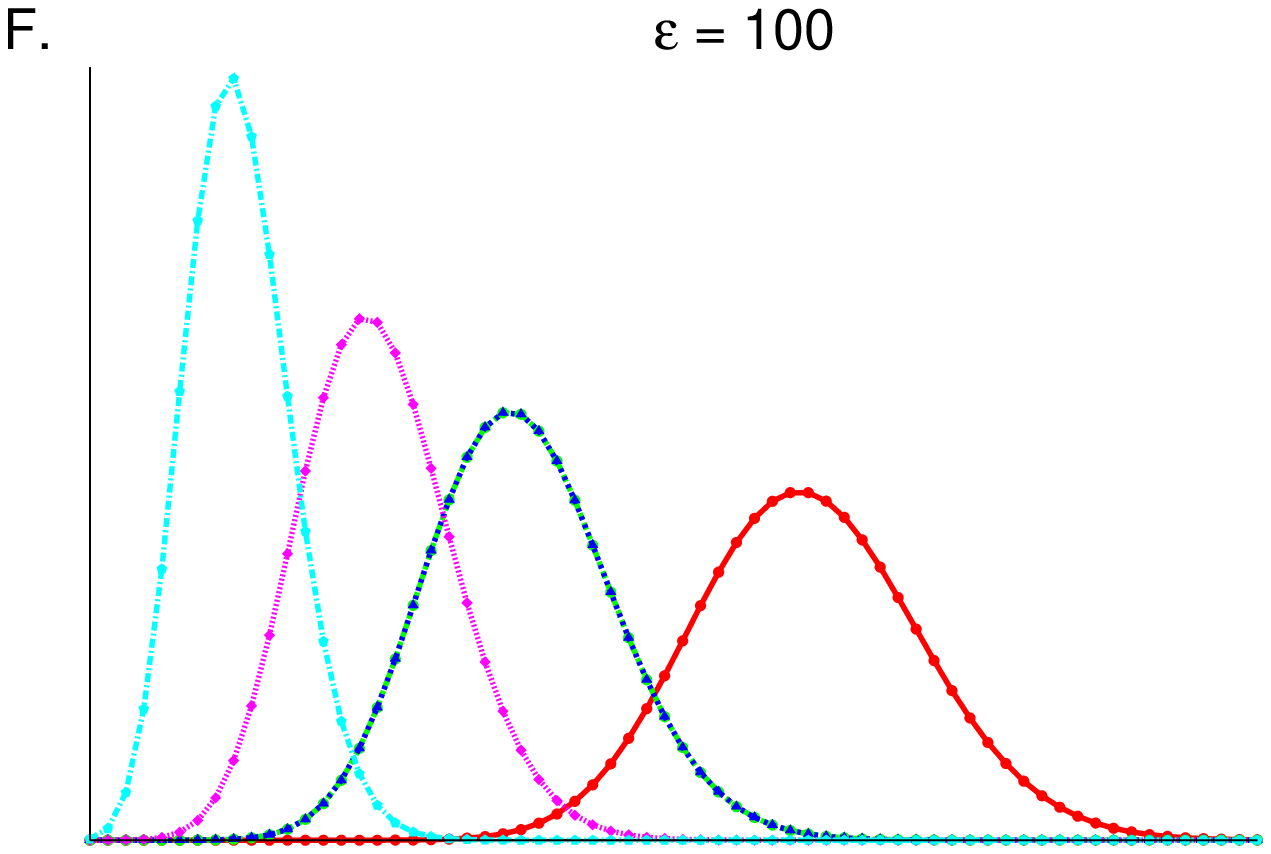}
       \caption{Efficient transcription,N=40. $\chi$ is chosen
       1/5. The values of $\epsilon$ and $p_\beta$ are shown in the figure.}
       \end{figure} 

       \newpage
      \begin{figure}[!htb]
       \includegraphics[width=2.0in,angle=0]{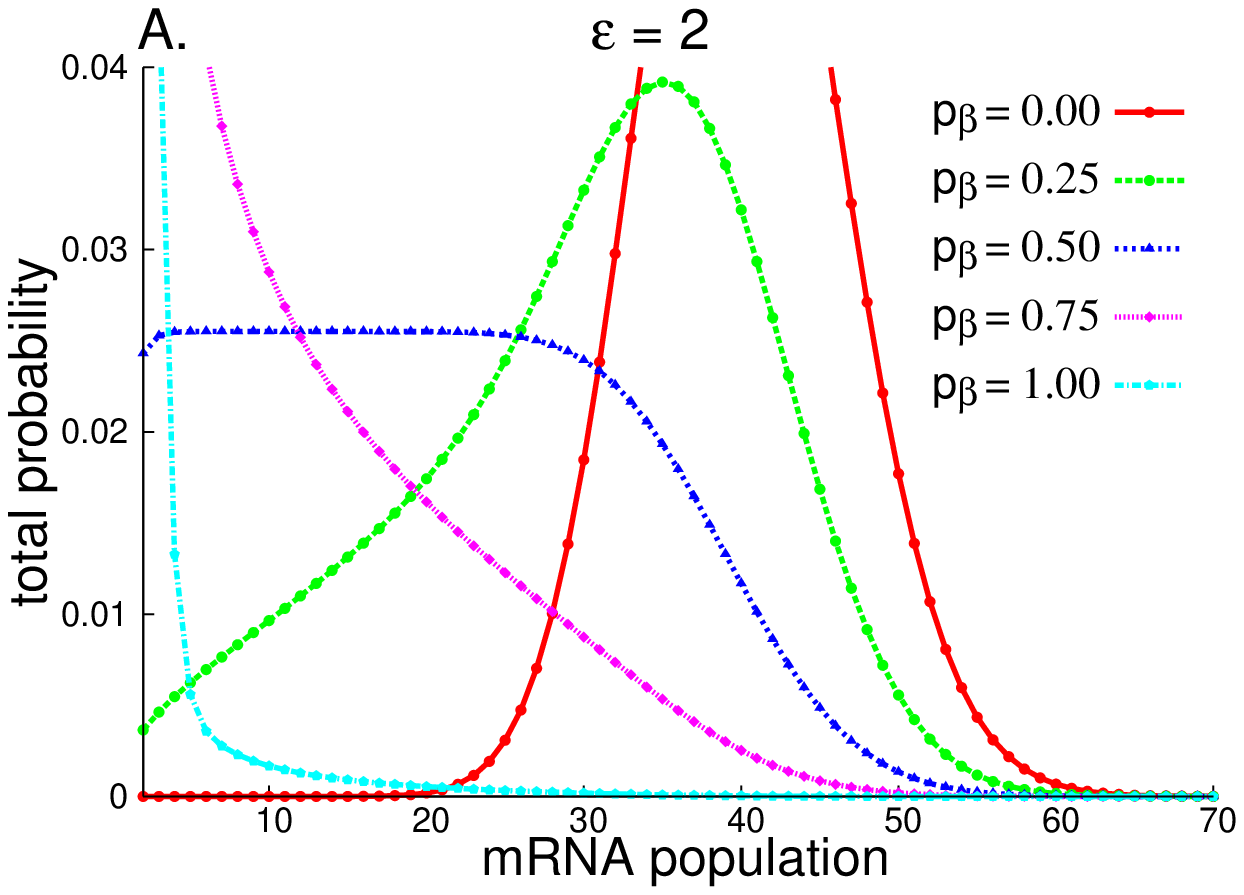}
       \includegraphics[width=2.0in,angle=0]{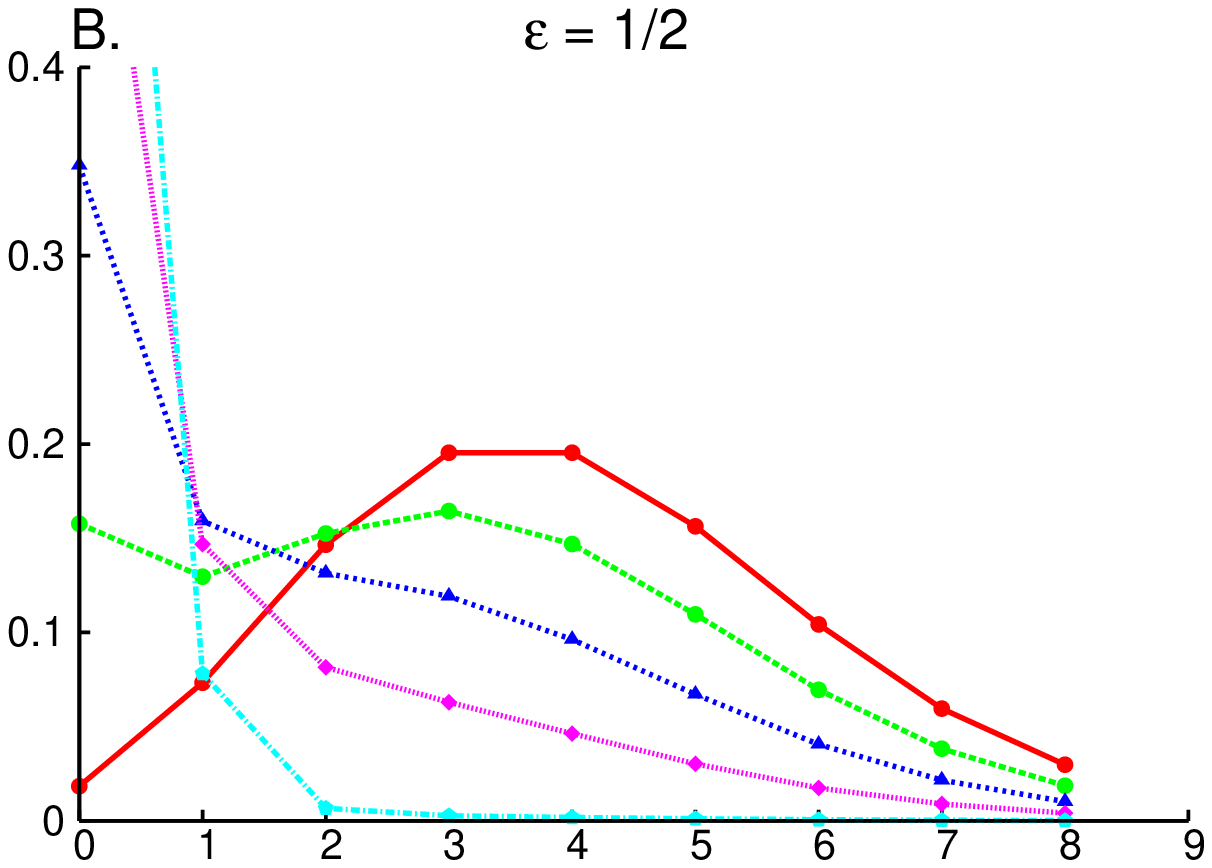}
       \includegraphics[width=2.0in,angle=0]{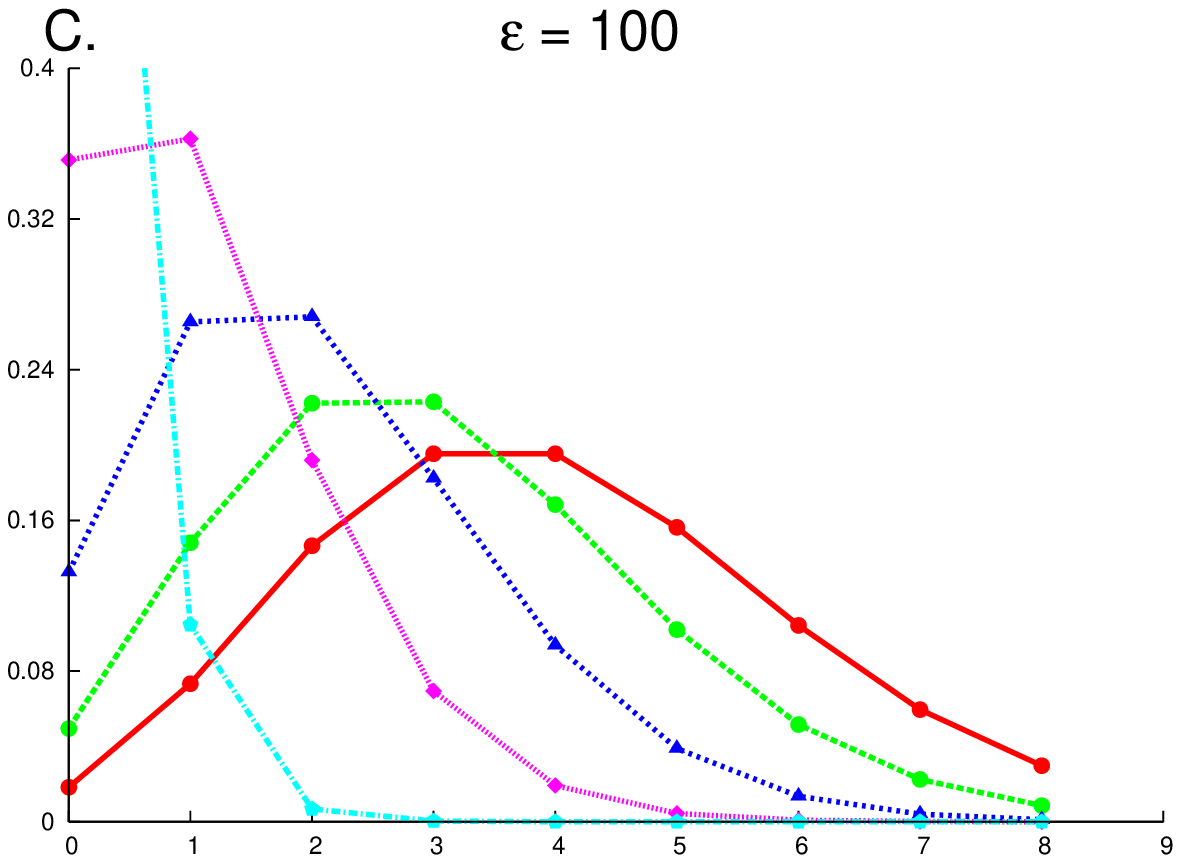}
       \caption{In Fig.4A $\epsilon$=2 and N=40.Figures 4B and 4C, N=4
        and $\epsilon$=1/2,100, respectively.$\chi$=1/50 in all cases}
       \end{figure} 
  
       \newpage
       \begin{figure}[!htb]
       \includegraphics[width=2.0in,angle=0]{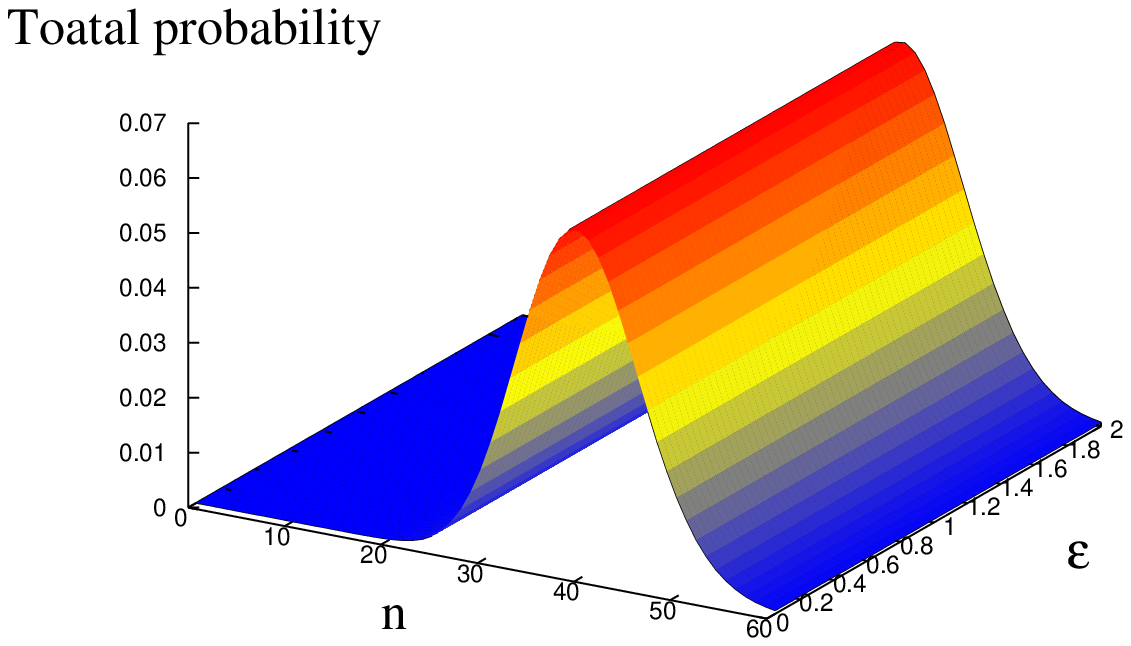}
       \includegraphics[width=2.0in,angle=0]{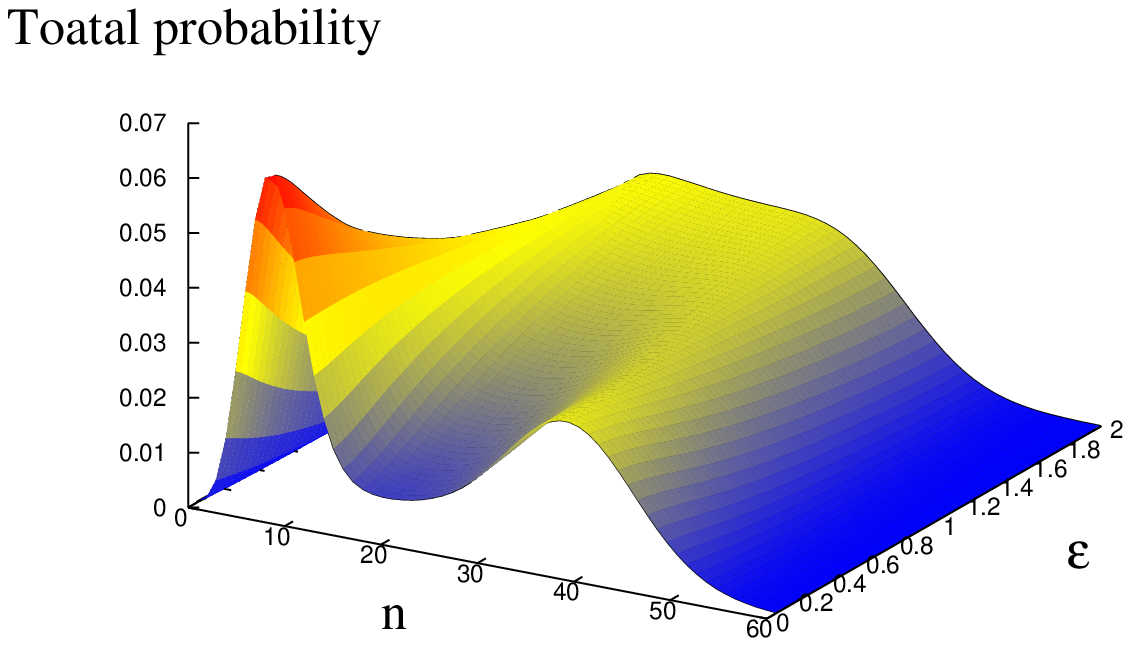}
       \includegraphics[width=2.0in,angle=0]{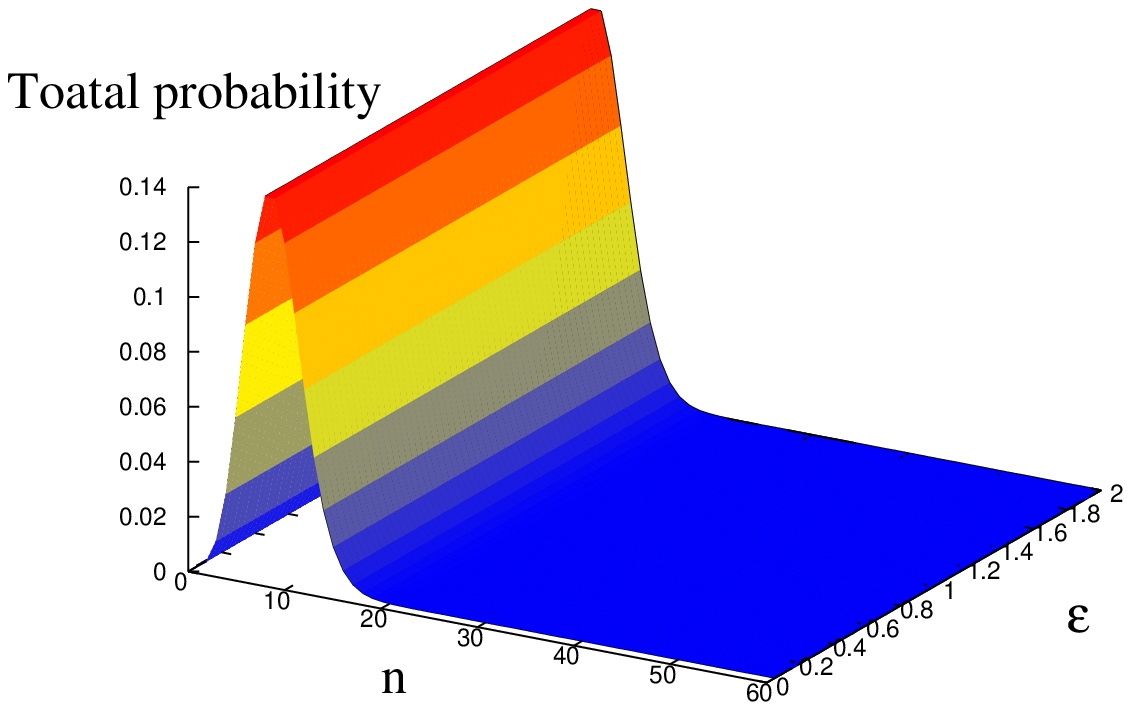}
       \caption{Surfaces for N=40 and $\chi$=1/5. In, A, B and C
       $p_\beta$=0, 1/2 and 1, respectively }
       \end{figure}

      \newpage 
       
      \begin{figure}[!htb]
      \centering
      \begin{minipage}[b]{0.65\linewidth}
      \includegraphics[width=\linewidth,angle=0]{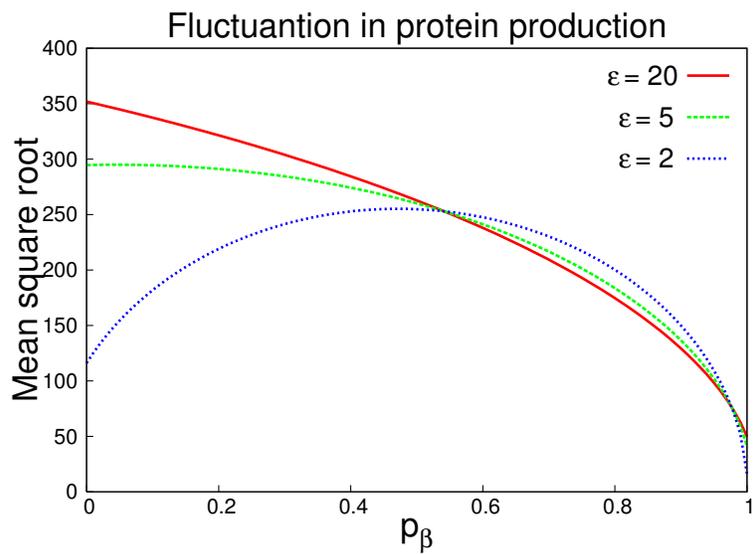}
      \end{minipage}
      \caption{Standard deviation in protein
        concentration. For all curves $\eta=1/20$ and $\chi=1/50$. The
        curve with $\epsilon=20$ the parameters are: $K_P=39.60$ and
        $K_R=4.04$. The curve with $\epsilon=5$, $K_P=27.44$ and
        $K_R=5.83$ and the curve with $\epsilon=2$, $K_P=3.37$ and
        $K_R=47.48$.}
      \end{figure}

      \newpage 
       
      \begin{figure}[!htb]
      \includegraphics[width=3in, angle=0]{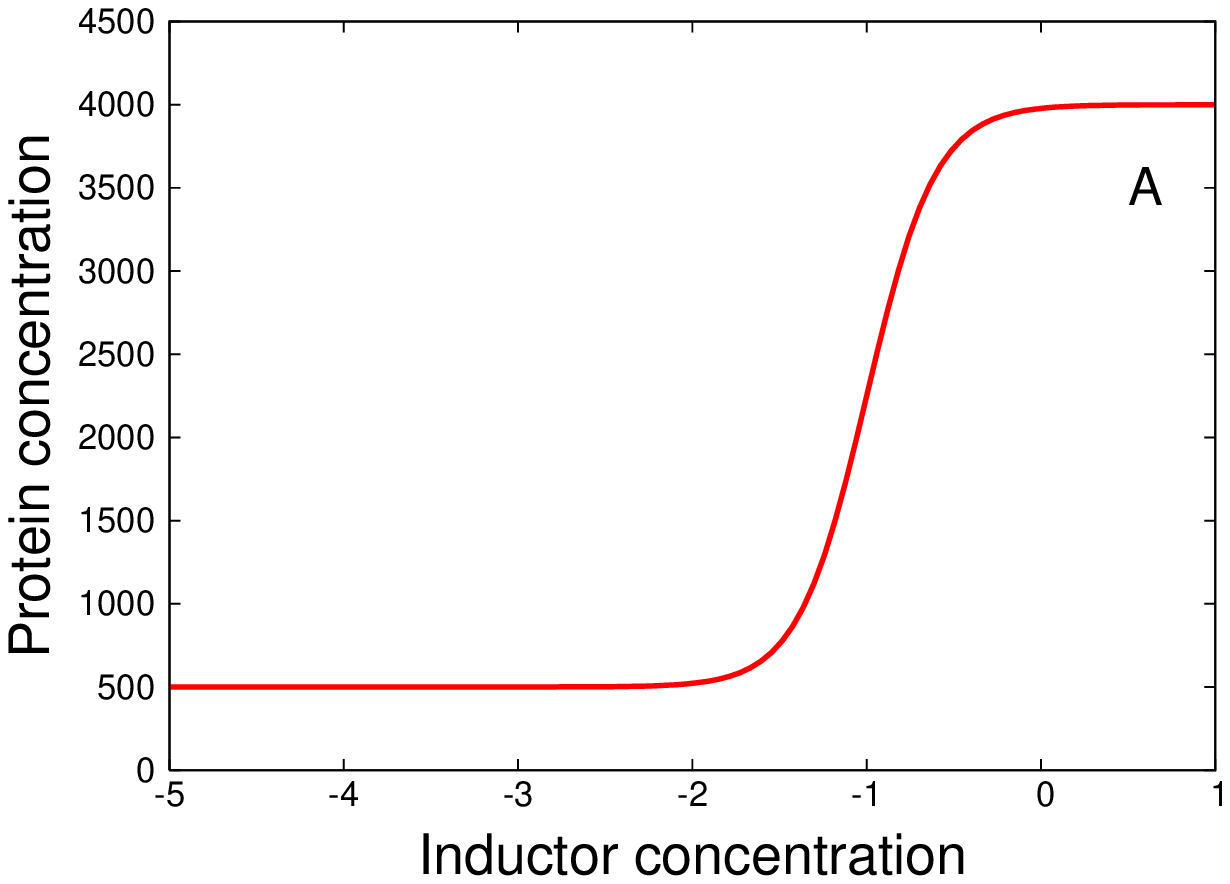}
      \includegraphics[width=3in, angle=0]{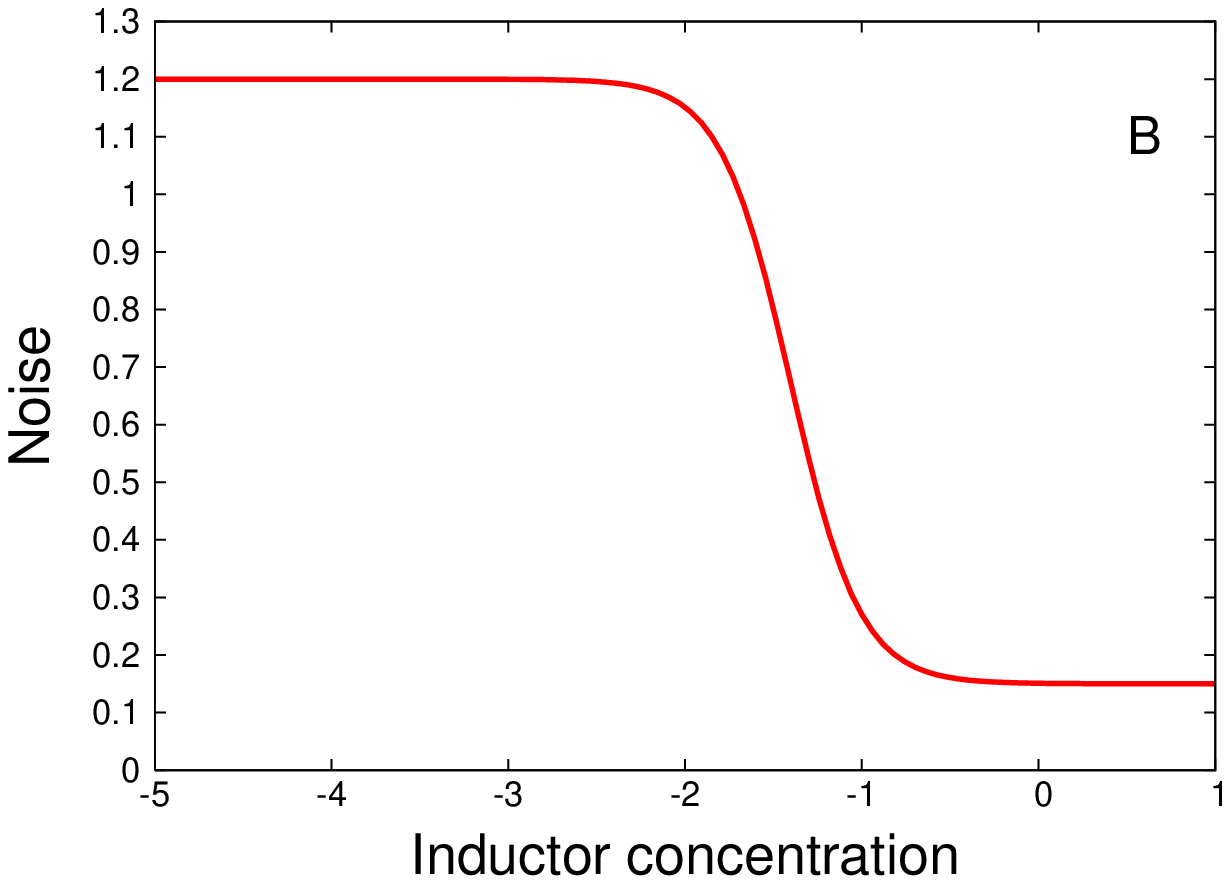}
     \caption{In figure 7.A and 7.B  the parameters are: $\sigma=730$, $N=1.20$
      $\eta=0.218$, $\chi=1/8$, $\epsilon=30$, $k=1/10$ and
      $\theta=2.2$. In  figure 7.B noise gene by equation (44) as a
      function  of the inductor concentration.}
      \end{figure}

      \newpage 
       
      \begin{figure}[!htb]
      \includegraphics[width=3in ,angle=0]{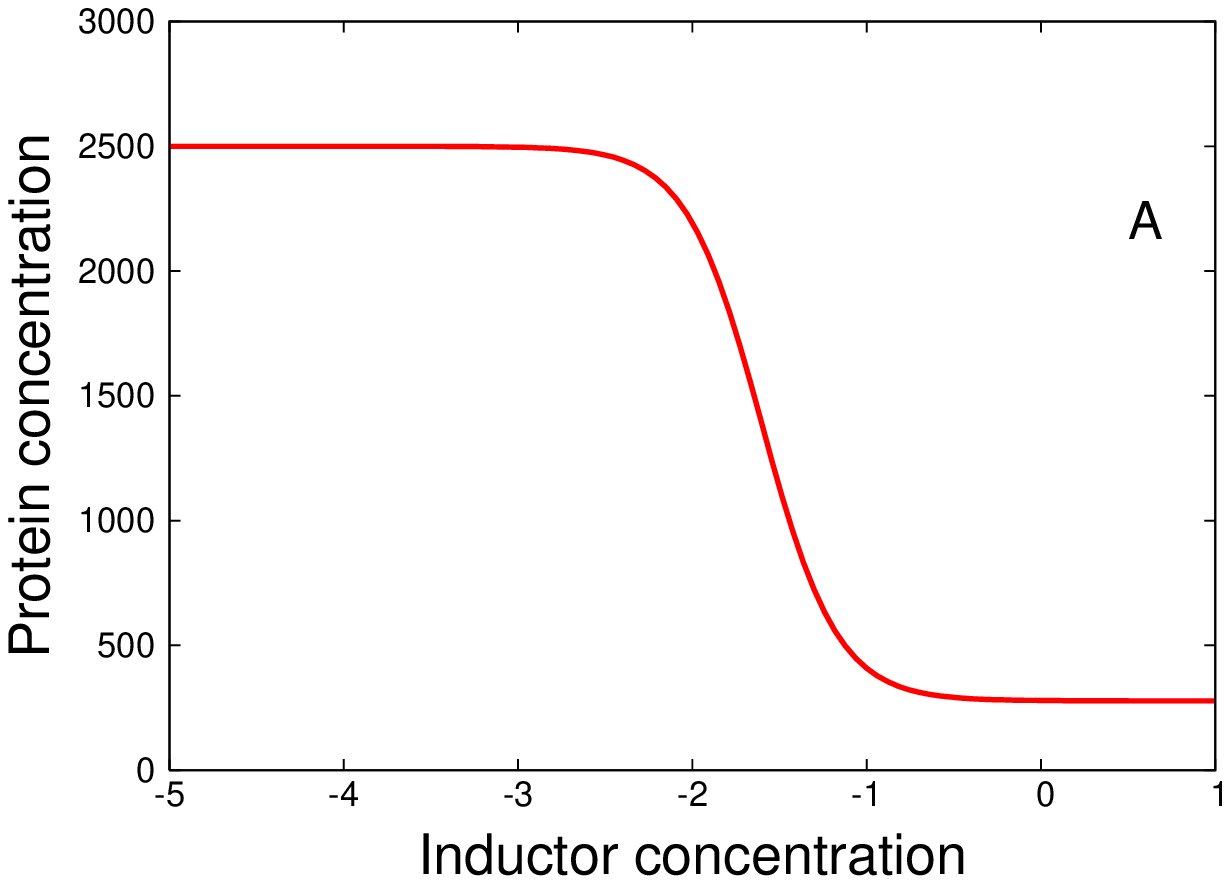}
      \includegraphics[width=3in ,angle=0]{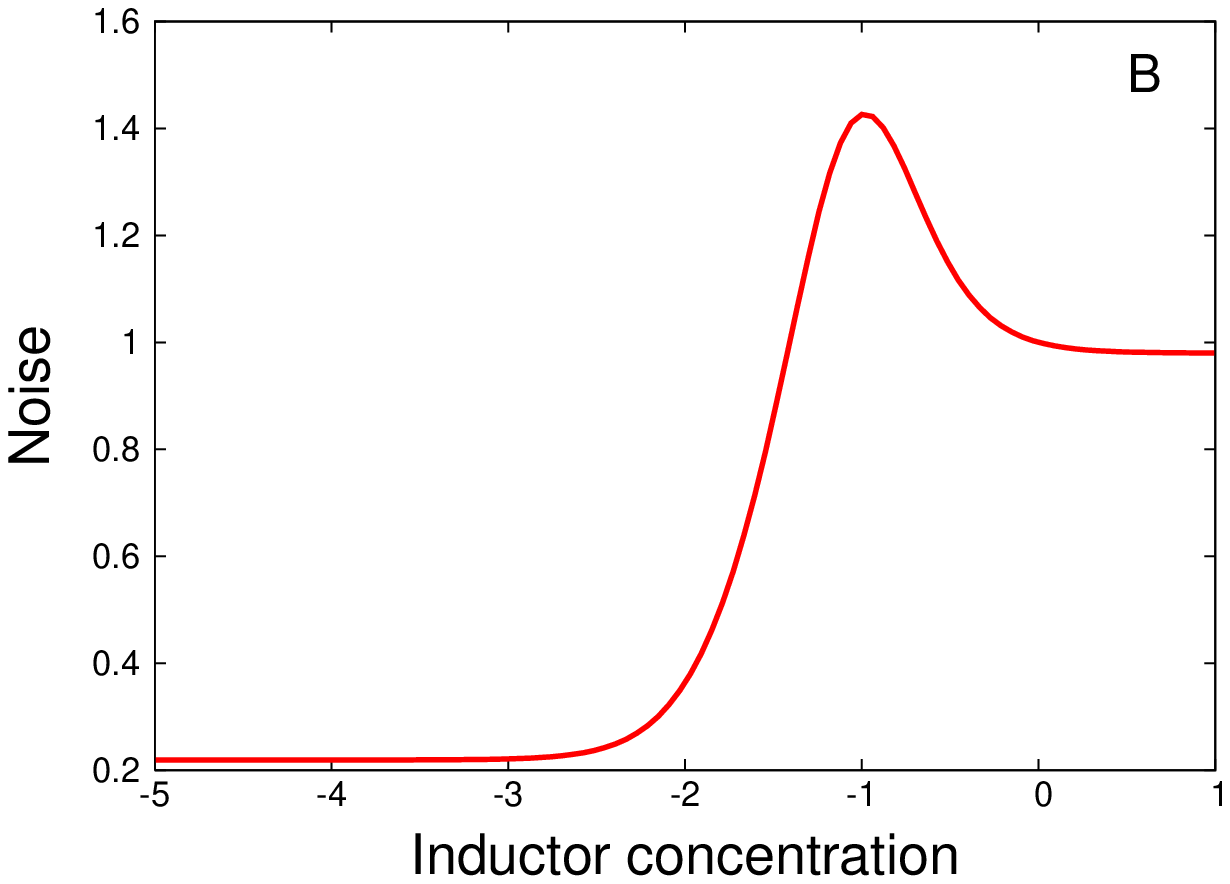}
      \caption{In both figures the parameters are: $\sigma=125$, $N=1$
      $\eta=1/20$, $\chi=1/20$, $\epsilon=1/10$, $k=1/40$ and
      $\theta=2$. In  figure 8.B noise gene by equation (44) as a
      function  of the inductor concentration.}
      \end{figure}

      \end{document}